# A Novel Approach for Video Temporal Annotation

S. Hadi Restgou Haghi [a1], Mohammadreza Kangavari [a], Behrang QasemiZadeh[a]

[a] *Iran University of Science and Technology, Computer Faculty, Narmak, Tehran, Iran*

Recent advances in computing, communication, and data storage have led to an increasing number of large digital libraries publicly available on the Internet. Main problem of content-based video retrieval is inferring semantics from raw video data. Video data play an important role in these libraries. Instead of words, a video retrieval system deals with collections of video records. Therefore, the system is confronted with the problem of video understanding. Because machine understanding of the video data is still an unsolved research problem, text annotations are usually used to describe the content of video data according to the annotator's understanding and the purpose of that video data. Most of proposed systems for video annotation are domain dependent. In addition, in many of these systems, an important feature of video data, temporality, is disregarded. In this paper, we proposed a framework for video temporal annotation. The proposed system uses domain knowledge and a time ontology to perform temporal annotation of input video.

## 1 INTRODUCTION

Recent advances in computing, communication, and data storage have led to an increasing number of large digital libraries publicly available on the Internet. In addition to alphanumeric data, other modalities, including video play an important role in these libraries. Ordinary techniques will not retrieve required information from the enormous mass of data stored in digital video libraries. Instead of words, a video retrieval system deals with collections of video records. Therefore, the system is confronted with the problem of video understanding. The system gathers key information from a video in order to allow users to query semantics instead of raw video data or video features. Users expect tools that automatically understand and manipulate the video content in the same structured way as a traditional database manages numeric and textual data. Consequently, content-based search and retrieval of video data becomes a challenging and important problem.

While the uses of content-based video retrieval are many, it has turned out to be a challenging problem. The difficulty lies in coming up with appropriate representations for the visual content which reflects the semantics of the video. Because machine understanding of the video data is still an unsolved research problem, text annotations are usually used to describe the content of video data according to the annotator's understanding and the purpose of that video data [1]. In general, computer vision techniques may aid in answering the question "what is in the video?" but cannot answer questions such as "what is happening in the video?" or "what is the video trying to tell us?"

The works on this era can be classified in two categories. At the first category, there has been increasing research effort put toward automatic generation of links between low-level features and high-level concepts, on the other words video data annotation to bridge semantic gap [2] while works on the other category emphasis on video modeling for efficient video data indexing and retrieval. Authors believe that works in both categories are in interaction with each other and best results can be achieved with an intelligent harmony between video content modeling and video annotation.

One of the principal things that makes video unique is that it is a temporal medium. Any language for annotating the content of video must have a way of talking about temporal events [3]. Any approach used for video modeling and abstraction should prepare a method for temporal information representation. Therefore, it is required that video annotations are accompanied with temporal information. This fact is (remains) unconsidered in many of previous works. The previous works, even they prepare a language for temporal information representation, do not consider about time and ontological issues about this topic in their framework.

In this paper, we have proposed a system for temporal annotation of video. The proposed system is then used as a part of framework which is introduced in [4]. The proposed system uses domain knowledge and a time ontology to provide temporal annotations of input video data. Later, these annotations will be used for video modeling as well as temporal reasoning on video evens and semantics. User defines the domain knowledge and time ontology according to the specific application. The rest of paper is organized as follows: section 2 gives a brief description of related works. The proposed system is described in section 3.

---
[1] Corresponding Author: Seyyed Hadi Rastgou Haghi, hadi.rastgou@gmail.com.

Conclusion and future works are discussed in section 4.

## 2 RELATED WORKS

Indexing and annotation systems for digital video files have been developed in the past - but only for use within stand-alone environments [5]. In [6], MPEG motion vectors, playfield shape and players position have been used with Hidden Markov Models to detect soccer highlights. In [7], a hierarchy of ontologies has been defined for the representation, of the results of video segmentation. Concepts are expressed in keywords and are mapped in an object ontology, a shot ontology and a semantic ontology. However, in this work there is no concern about temporal information of underlying video.

In [2], a generic system for automatic annotation of videos is introduced. In the proposed system, knowledge embedded in the video database and interaction with an expert user is exploited to enable system learning and generate a rule-knowledge base to annotate video data. Same as other works, here also there is no concern about temporality of video data and temporal relations between annotations. Other examples of such systems can be found in [13] and [15].

As we mentioned, in most of these works, proposed frameworks do not support temporal information representation as a part of their annotations. Even supporting temporal information in these frameworks, such systems have an important short comes about ontological issues about time. Actually, in many of these works, there is no concern about explicit description of time and its abstraction in applications. These systems do not specify how these annotations can be used for temporal reasoning about their underlying videos. Also most of these frameworks are domain specific and they do not show how they can be used for different kind of videos.

## 3 THE PROPOSED SYSTEM

Figure 1 shows the architecture of the proposed system. The system is designed as a part of larger system which has been introduced in [4]. The input of the proposed system is raw video data. the output of system is temporal annotation according to the input video data. Later, these annotations will be used for video data modeling. The system comprises of a time ontology, domain knowledge, user, meta-rule editor, video analysis module, and temporal video annotator.

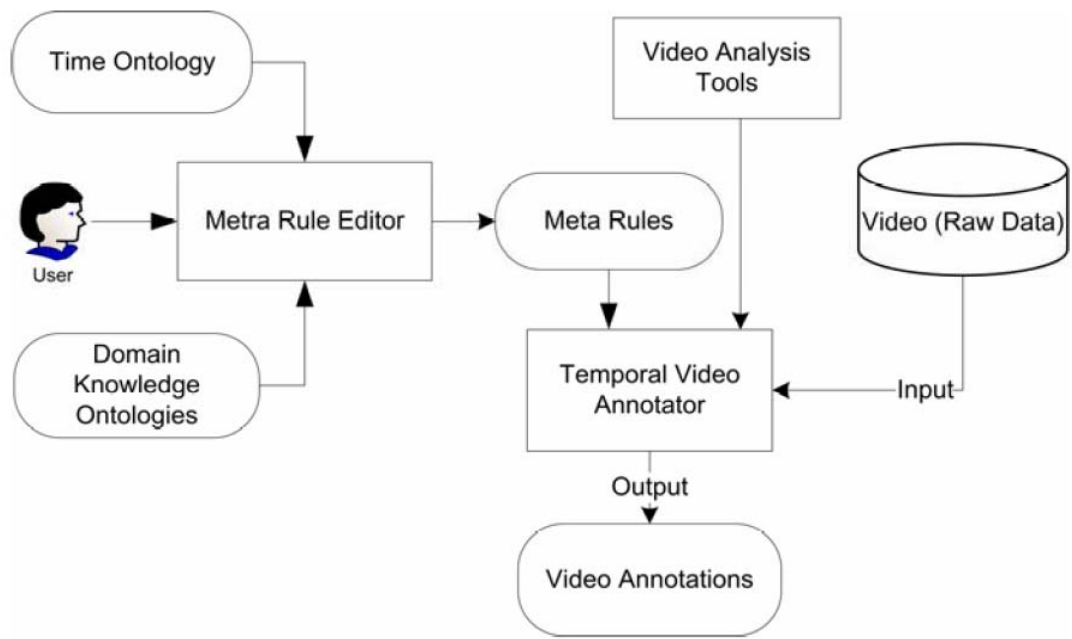

Fig. 1. Block Diagram of the Proposed System

In the proposed system, domain knowledge module represents knowledge of domain application in terms of ontology. User may define the domain ontology using prepared Graphical User Interface (GUI); or he/she can import it from other predefined ontologies. The ontology contains general rules, objects, concepts, etc. of interest in a given context. For example, if the system is used to annotate soccer matches, then the domain ontology contains facts about objects such as players, teams, etc; also it is possible that the concepts like win, lose, fault and etc. are defined in domain ontology. Further more, each entity in domain ontology could be related to temporal entity in time ontology.

Time ontology comprises of facts about time. In fact, time ontology indicates the underlying abstraction of time in a domain application. All the ontological issues about time and temporal reasoning are defined in time ontology. As an example, in a soccer match half time is defined 45 minutes, while in basketball half time is defined 20 minutes. In addition, different relations between temporal entities, such as before, after, et cetera are defined in time ontology. More information about time ontology and intended meaning can be found in [4].

Video analysis module provides a set of tools for video analysis such as shot detection, text segmentation, text tracking and so on. These services then can be used by user to define meta-rules to extract proper annotations. As the current version, we have used a set of tools which are obtainable from MoCA project [8], [9], and [10].

Meta-rule editor provides a GUI to define meta-rules. Figure 2 shows the GUI. The GUI offers a set of logical operations as well arithmetic one, video analysis tools, entities of domain knowledge like keywords, concepts and the rest for user to define meta-rules. In addition, user may use time ontology and its axioms in definition of rules. Note that according to the domain ontology, time ontology and prepared video analysis tools, meta-rules can vary from one application to the other application. In this way, the proposed system can be used independent of domain. Note that like domain knowledge, it is possible to load previously defined meta-rules. A simple meta-rule could be composed of text tracking, text segmentation, OCR, and template matching to extract information from a part of video.

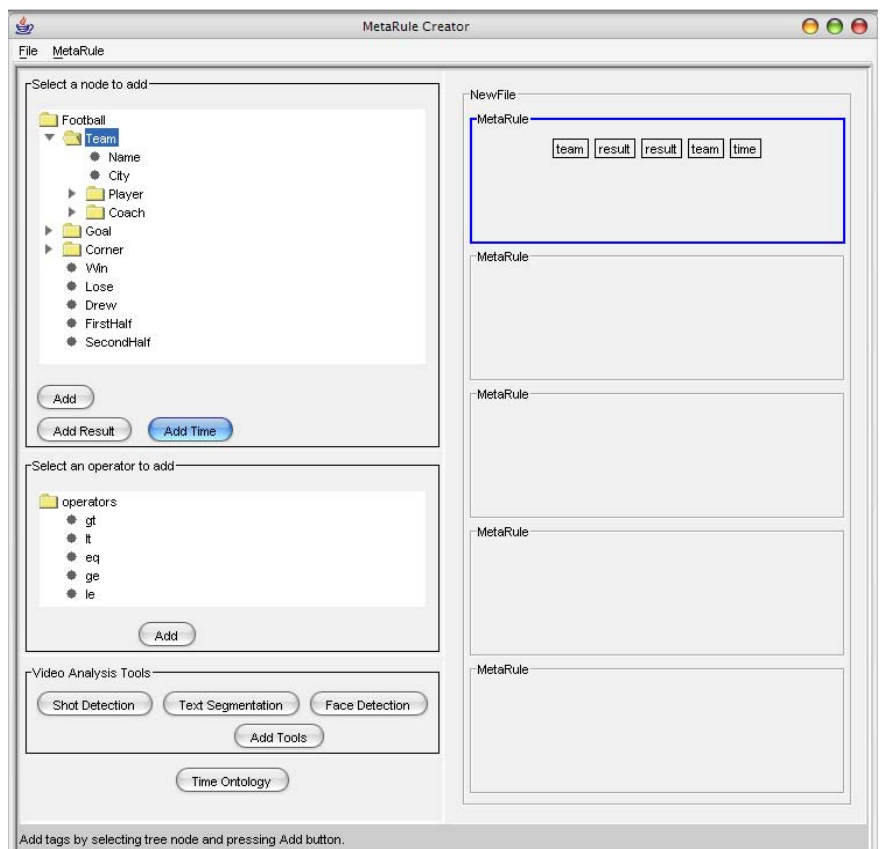

Fig. 2. Graphical User Interface of meta-rule Editor

Temporal video annotator uses meta-rules to annotate input video data. When user defines meta-rules for the system, temporal video annotator uses these meta-rules to extract proper annotations from raw input video data. The temporal video annotator parses predefined meta-rules and it performs suitable operations, functions as well as inferences to make each part of video related to a part of domain knowledge and time ontology. In actual words, the annotations are formulated using relations between video segments and ontologies to deliver high-level concepts from raw video data to the user. The annotations further could be used for video abstraction and modelling.

## 4 CONCLUSION AND FUTURE WORKS

Given the current state of the art in machine vision and image processing, it is not possible that machines "watch" and understand the content of digital video archives for us. Although some aspects of video can be automatically parsed, a detailed representation of video content requires that video be annotated. Therefore, the central problem in manipulating video information lies in representing and visualizing video content. Since now, several annotation systems have been developed but only for use within specific application. In addition, in many of these systems there is no concern about an important feature of video documents, temporality.

This paper introduced a novel approach for temporal annotation of video data, which is domain independent. The system generates uses domain ontologies as well as a time ontology to generate suitable temporal annotations. To achieve this goal, user must define meta-rules from the elements of domain ontology, time ontology, video analysis tools, and arithmetic and logical operators. Then, system uses these meta-rules to prepare appropriate annotations with temporal information. The generated annotations then could be used for video data modeling as shown in [4].

One of the Advantages of the system is that it can be used for different application. Because domain knowledge is introduced to the system in terms of ontologies, user can define or import these ontologies according to his/her interest and application. Moreover, user can change the abstraction about time and temporal issues by defining suitable time ontologies which is not concerned in prior works.